%% file: 0_main.tex
\documentclass[fontsize=11pt]{scrartcl}
\usepackage[margin=1in]{geometry}
\usepackage{kpfonts}
\newcommand\mnras{Mon. Not. Roy. Astron. Soc}
\newcommand\apjs{The Astrophysical Journal Supplement Series}
\newcommand\apj{The Astrophysical Journal}
\newcommand\na{New Astronomy}
\usepackage{amsmath}
\usepackage{textcomp}
\usepackage{cite}
\usepackage{graphicx}
\usepackage[font={footnotesize}]{caption}
\usepackage{subcaption}
\usepackage[section]{placeins}
\usepackage{float}
\usepackage{subfiles}
\usepackage{enumitem}
\usepackage{afterpage}
\usepackage[T1]{fontenc} 
\usepackage[english]{babel} 
\usepackage[en-US]{datetime2}
\DTMlangsetup{showdayofmonth=false}
\usepackage{blindtext}
\usepackage{booktabs}
\usepackage{physics} 
\usepackage{sectsty}
\makeatletter
\newcommand\footnoteref[1]{\protected@xdef\@thefnmark{\ref{#1}}\@footnotemark}
\makeatother

\usepackage{hyperref}
\addto\extrasenglish{

}

\sectionfont{\fontsize{14}{14}\normalfont\bfseries}
\subsectionfont{\fontsize{14}{14}\normalfont\bfseries}
\usepackage{fancyhdr} 
\begin{document}


\begin{center}
$\>\>$\\
\DTMlangsetup{showdayofmonth=true}
\Large \textbf{Lightcone Construction for HACC Cosmological Simulations with LANTERN}\\
\normalsize Joseph Hollowed\\Cosmological Physics and Advanced Computing Group, Argonne National Laboratory\\
\textit{Last edited June 18, 2019}
\end{center}
$\>\>$\\
This document describes the LANTERN (Lightcone generAtion via sNapshoT intERpolatioN) code module, our current approach for generating lightcones from HACC \cite{2016NewA...42...49H} simulation products. In \autoref{sec:deriv}, we derive the condition which defines all spacetime events that can be seen by an observer at any given time (we parameterize the surface of the observer's past lightcone). In \autoref{sec:cross}, we describe how we apply this condition to find approximate lightcone-crossing times of objects whose positions and velocities are known only at discrete time snapshots. Higher order effects and corrections are discussed in \autoref{sec:nonlin}. \autoref{sec:geom} describes simulation box replication capabilities that allow generation of arbitrary lightcone sizes.


\section{Parameterizing the Lightcone Surface}\label{sec:deriv}

\subfile{1_derivation}

\section{Crossing the Lightcone}\label{sec:cross}

\subfile{2.0_crossing}

\subsection{Particle Extrapolation}\label{sec:extra}

\subfile{2.1_extrapolation}

\subsection{Particle Duplication}\label{sec:dupl}

\subfile{2.2_duplication}

\subsection{Particle Interpolation}\label{sec:inter}

\subfile{2.3_interpolation}

\section{Nonlinear Effects}\label{sec:nonlin}

\subfile{3_nonlinear}

\section{Lightcone Depths Beyond the Simulation Volume}\label{sec:geom}

\subfile{4.2_geometry}

\section{Code Modularity}\label{sec:export}
\subfile{5_export}

\section{Acknowledgements}\label{sec:credits}

\subfile{6_credits}

\pagebreak
\subfile{7_bib}
\bibliographystyle{plain}

\end{document}

%% file: 1_derivation.tex
At any given time $t$, an observer will see only what is on the surface of their past-lightcone. Any event located inside the lightcone surface has been seen in the past, and any event outside will be seen in the future, in general. Hence, our goal is to parameterize this surface in terms of quantities that are available to us as simulation outputs. Applying this parameterization to simulation datasets will allow us to locate all particles on an observer's past-lightcone at any given time, yielding the observable universe.

Let us first consider the problem in a static Euclidean space. Recall that the spacetime separation between an observation, occurring at $(t, r, \theta, \phi)$, and some event at $(t+dt, r+dr, \theta+d\theta, \phi+d\phi)$ is given by the Minkowski metric:
\begin{equation}
    ds^2 = -c^2dt^2 + dr^2 + r^2d\Omega^2 ,
\end{equation}
where $d\Omega^2$ includes the change of variables from Cartesian to spherical coordinates as
\begin{equation}
    d\Omega^2 \equiv d\theta^2 + \sin^2\theta d\phi^2.
\end{equation}
An event $A$ being located on the lightcone surface is defined by the fact that the path joining it to the observer is a null geodesic, where $ds^2 = 0$. If we impose the constraint that the observer is located at the spatial origin, then this path is constant in $\theta$ and $\phi$, and thus $d\Omega^2 = 0$. Objects located on the lightcone therefore satisfy the condition
\begin{equation}
        0 = -c^2dt^2 + dr^2
\end{equation}
which can be rearranged to express the distance to $A$ in terms of the light travel time as 
\begin{equation}
    c^2dt^2 = dr^2 \>\>\> \rightarrow \>\>\> \int_{t_{\text{em}}}^{t_{\text{obs}}} 
    c\>dt = c(t_\text{obs} - t_\text{em}) = r,
\end{equation}
where $t_\text{obs}$ and $t_\text{em}$ are the time coordinates of the observer and event $A$ (signal observation and emission), respectively. $r$ is the magnitude of the spatial separation between the observer and $A$.

So, we have parameterized the lightcone surface by the rightmost equality in Eq.(4) for a static universe. However, the universe \textit{is not} static, and so light travel time between two events is \textit{not} $r/c$, but should effectively be modified by the scale factor $a(t)$. Obtaining the correct condition for finding an event on the lightcone, then, must require repeating the simple procedure in Eq.(1-4) using the more general Robertson-Walker metric:
\begin{equation}
    ds^2 = -c^2dt^2 + a(t)^2\left[ dr^2 + S_\kappa(r)^2d\Omega^2\right].
\end{equation}
Here, $S_\kappa(r)$ defines the curvature of space, which can be ignored since $d\Omega^2=0$, as per our assumptions stated above (the curvature information remains encoded in the form of $a(t)$). In that case, a null spacetime separation under this metric gives
\begin{equation}
    \int_{t_\text{em}}^{t_\text{obs}}c\dfrac{dt}{a(t)} = r.
\end{equation}
It will also be handy to have this condition in another form; we can use the Hubble parameter $H$ to cast the variable of integration to the scale factor, as a proxy for time, if we note the following substitutions:
\begin{equation}
    dt = \dfrac{dt}{da}da = \dfrac{da}{\dot{a}} \>\>\>\>\>\>\> \text{and} \>\>\>\>\>\>\>
    H = \dfrac{\dot{a}}{a}\>.
\end{equation}
Combining Eq.(6-7) results in
\begin{equation}
    \int_{a_\text{em}}^1 c\dfrac{da}{a^2H} = r.
\end{equation}
Eq.(6, 8) is our sought after parameterization\footnote{Eq.(8) is actually the parameterization of a sphere in three-dimensional space, whose radius is dependent on the parameter $a$. Actually visualizing such a sphere on a static figure requires projecting the object into two spatial dimensions and allowing time to occupy the third axis, visually resulting in a cone.}; all events whose spacetime coordinates satisfy this condition are found on the lightcone surface and are visible to the observer, at the spatial origin, at time $t_{\text{obs}}$.

%% file: 2.0_crossing.tex
Given pristine knowledge of particle positions resultant from cosmological simulations, Eq.(8) appears straightforward to apply: for any time $t$, find the associated scale factor $a(t)$, and evaluate the integral on the left hand side of Eq.(8) to obtain some value $R$. Then, check all particles in the simulation for the condition that $r_i=R$, where the subscript $i$ denotes the $i$th particle. If the condition is satisfied for any particle $i$, then that particle crossed the lightcone and is seen by the observer at time $t$. Repeat this for all times $t$ to fill the lightcone.

Unfortunately, simulation outputs do not include ``all times $t$". Since we only have of order $100$ time snapshot outputs from $a\approx0$ to $a=1$, the naive approach described in the previous paragraph is not applicable. Consider a particle $i$ that crosses the lightcone surface (is seen) at time $t_\text{em}$\footnote{To say that the particle ``is seen" at time $t_\text{em}$ is shorthand for the more accurate statement: ``photons emitted from the particle's position at time $t_\text{em}$ intersect the observer's position at time $t_\text{obs}$". The observer of course observes the entire sky at a single time $t_\text{obs}$.}. This event is captured in the simulation outputs only by the fact that particle $i$'s separation with respect to the observer is timelike ($ds^2<0$) at some snapshot time $t_j < t_\text{em}$, and spacelike ($ds^2>0$) at the next snapshot time $t_{j+1} > t_\text{em}$. Thus, we need some prescription for approximating particle trajectories between these times, to find $t_\text{em}$ and $r$ (where $ds^2=0$). 

Several such methods have been described in the literature; some have chosen to pull unaltered particle positions from simulation snapshots, and assign redshifts analytically \cite{Blaizot:2003av, Kitzbichler:2006ec}, while others interpolate particle positions between snapshots using linear or higher order methods \cite{evrard2002, 2013MNRAS.429..556M, Smith:2017tzz}. These decisions are made with the desired characteristics of the output in mind-- using snapshot positions, for instance, preserves the smallest scale features of the base N-body run, though will introduce discontinuities in redshift dependent statistics as the lightcone content jumps from one snapshot to the next (and vice-verse for approximating the unknown inter-snapshot particle trajectories)

In the HACC codebase, particle position extrapolation has traditionally been used, from time $t_j$, which is discussed in \autoref{sec:extra}. More recently implemented is LANTERN, an improved approach which rather linearly interpolates between snapshot times $t_j$ and $t_{j+1}$, described in 
\autoref{sec:inter}.

%% file: 2.1_extrapolation.tex
Since we would ultimately like to solve Eq.(8) for $a_\text{em}$ (equivalently Eq.(6) for $t_\text{em}$), we must first write the relevant quantities in terms of simulation data products. Let us consider a single simulation timestep for the remainder of this section, which begins at snapshot $j$ and ends at snapshot $j+1$\footnote{This notation is not meant to imply that lightcone construction is necessarily performed on \textit{all} simulation timesteps. $j$ and $j+1$ are only meant to refer to adjacent simulation snapshots in whatever output set is in use.}. The time it spans is $t_{j+1} - t_j$, and the scale factor evolution is $a_{j+1} - a_j$. 

\begin{figure}[t!]
  \begin{minipage}[c]{0.65\textwidth}
    \includegraphics[width=\textwidth]{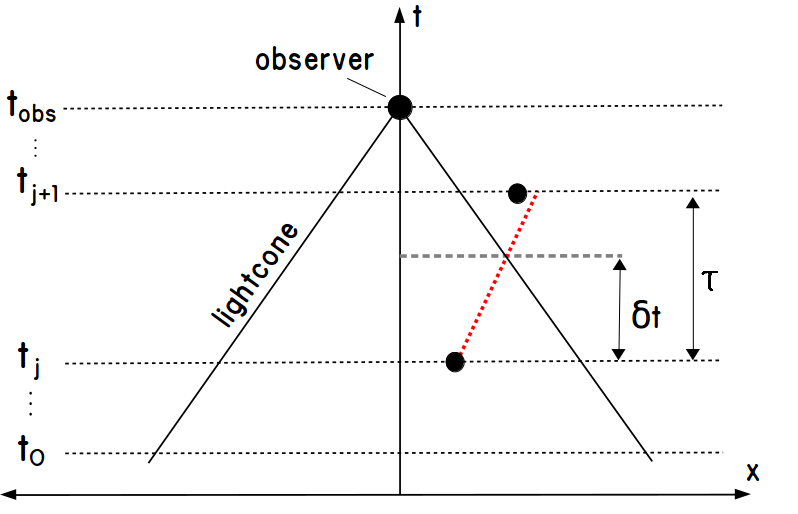}
  \end{minipage}\hfill
  \begin{minipage}[c]{0.35\textwidth}
    \caption{
    \textbf{Figure 1:} A schematic of linear particle position extrapolation. This figure focuses on an arbitrary timestep bounded by snapshots $j$ and $j+1$, where the vertical axis is time and the horizontal axis is space. Time increases upward. The particle (black disc) is found inside the lightcone at time $t_j$, and outside the lightcone at the time $t_{j+1}$ of the following snapshot. It is estimated to cross the lightcone, by extrapolation (red dashed line), at time $t_j + \delta t$. Note that the extrapolation does not in general estimate the correct position of the particle at time $t_{j+1}$.
    } \label{fig:1}
  \end{minipage}
\end{figure}

Each particle in snapshot $j$ has six quantities that are relevant to our purposes; three comoving Cartesian coordinates and corresponding velocities 
\begin{equation}
    \textbf{r}_j = \begin{bmatrix} x_j \\ y_j \\ z_j \end{bmatrix} \>\>,
        \>\>\>\>\>\>\> \textbf{v}_j = \dot{\textbf{r}}_j\>\>.
\end{equation}
We now approximate $r$ by linear extrapolation from the particle's position and velocity at $t_j$:
\begin{align}
    r &\approx \|\textbf{r}_j + \textbf{v}_j\delta t\|\nonumber\\
    &= \left[(x_j + \dot{x}_j\delta t)^2 + (y_1 + \dot{y}_j\delta t)^2 + (z_1 + \dot{z}_j\delta t)^2\right]^{1/2}
\end{align}
where we have introduced the quantity $\delta t$ as $\delta t = t_\text{em} - t_j$. Let us also obtain an expression for $a_\text{em}$ in Eq.(8), the scale factor at the time $t_j + \delta t$ to first order:
\begin{equation}
    a_\text{em} = a(t_j + \delta t) \approx a_j + \dot{a}_j\delta t\>,
\end{equation}
and describe $a_{j+1}$ in terms of $a_j$ and the timestep width $\tau = t_{j+1}-t_j$, as
\begin{equation}
    a_{j+1} \approx a_j + \dot{a}_j\tau \>.
\end{equation}
Refer to \autoref{fig:1} for a visual summary of the important quantities discussed above.

\captionsetup{labelformat=empty}
\captionsetup{labelformat=default}

Note that we have taken the assumption that $0 \leq \delta t \leq \tau \ll 1$, retaining up to second order terms in $\delta t$, and assuming that $\ddot{\textbf{r}}_j=0$\footnote{With certain assumptions made about our time units, the assertion that $\tau \ll 1$ can seem absurd; note that our temporal unit is \textit{program time}, which is the scale factor to a power $\alpha$, typically taken to be $\alpha=1$. Thus, this assumption is appropriate for all but very coarsely resolved time outputs.}. This allows us to further approximate $r$ as
\begin{align}
    \|\textbf{r}_\text{event}\| &= \|\textbf{r}_j + \textbf{v}_j\delta t\|\nonumber\\
    &\approx \|\textbf{r}_j\| + \delta t\dfrac{(\textbf{r}_j\cdot\textbf{v}_j)}{\|\textbf{r}_j\|} - \frac{\delta t^2}{2}\frac{(\textbf{r}_j\cdot\textbf{v}_j)^2}{\|\textbf{r}_j\|^3} + \frac{\delta t^2}{2}\frac{(\textbf{v}_j \cdot \textbf{v}_j)}{\|\textbf{r}_j\|}
\end{align}
in order to isolate $\delta t$\footnote{A subtle approximation made here; write it as $r^2 = (\textbf{r}_j+\textbf{v}_j\delta t)\cdot(\textbf{r}_j+\textbf{v}_j\delta t)$, expand, and apply the binomial series to second order, dropping terms higher than $\delta t^2$.}. Next, we can break the left-hand-side of Eq.(8) into two integrals; one that spans the time from the particle's lightcone intersection (at $a_\text{em}$) to the end of the snapshot $t_{j+1}$ (at $a_{j+1}$), and one that spans the remaining history of the universe:
\begin{equation}
    \int_{a_\text{em}}^1 c\dfrac{da}{a^2H} = \int_{a_\text{em}}^{a_{j+1}} c\dfrac{da}{a^2H} + \int_{a_{j+1}}^1 c\dfrac{da}{a^2H}\>.
\end{equation}
The first thing to note, here, is that the the bounds and integrand of the latter piece over $[a_{j+1},1]$ are entirely known in the simulation parameters and snapshot information, when we express $H$ as 
\begin{equation}
    H(t) = \dfrac{H_0}{a(t)^{3/2}}\sqrt{\Omega_m+\Omega_\Lambda a(t)^3}\>.
\end{equation}
So, we can calculate this integral numerically via Simpson's rule quadrature, and will henceforth refer to that result as 
\begin{equation}
    \Theta_{j+1} \equiv \text{Simpson}\left(\int_{a_{j+1}}^1 c\dfrac{da}{a^2H}\right)\>.
\end{equation}
$\>$

Turning our attention to the $[a_\text{em}, a_{j+1}]$ piece of Eq.(14), we see that we cannot trivially numerically evaluate this integral, since we do not know $a_\text{em}$--- let's instead solve the integral explicitly, approximating the result by dropping terms higher than $\delta t^2$. We write this piece in the form of Eq.(6) and change the variable of integration from $t$ to $t' = (t - t_j)$:
\begin{align}
    \int_{a_\text{em}}^{a_{j+1}}c\dfrac{da}{a^2H} &= \int_{t_\text{em}}^{t_{j+1}} c\frac{dt}{a(t)}
    \nonumber\\
	&\approx \int_{\delta t}^\tau c\frac{dt'}{a_j + \dot{a}_j(t')}\nonumber\\
    &\approx \frac{c}{a_j}\left[(\tau-\delta t) - \frac{\dot{a}_j}{a_j}\frac{(\tau^2-\delta t^2)}{2}\right]
\end{align}
With that, we have all of the ingredients required to cast Eq.(8) in terms of quantities available in our simulation output. The left hand side of Eq.(8) is replaced with Eq.(16-17), while the right hand side is replaced with Eq.(13):
\begin{equation}\tag{8}
    \int_{a_\text{em}}^1 c\dfrac{da}{a^2H} = r
\end{equation}
\begin{center}$\downarrow$\end{center}
\begin{align}
    \Theta_{j+1} + \frac{c}{a_j}\left[(\tau-\delta t) - \frac{\dot{a}_j}{a_j}\frac{(\tau^2-\delta t^2)}{2}\right] &= \|\textbf{r}_j\| + \delta t\dfrac{(\textbf{r}_j\cdot\textbf{v}_j)}{\|\textbf{r}_j\|} - \frac{\delta t^2}{2}\frac{(\textbf{r}_j\cdot\textbf{v}_j)^2}{\|\textbf{r}_j\|^3} + \frac{\delta t^2}{2}\frac{(\textbf{v}_j \cdot \textbf{v}_j)}{\|\textbf{r}_j\|}
\end{align}
$\>$\\
which we can then finally solve for $\delta t$. The most apparent way to do this is via a quadratic formula
\begin{align}
\alpha\cdot \delta t^2 + \beta\cdot \delta t + \gamma &= 0\\
\implies \delta t &= \dfrac{-\beta\pm\sqrt{\beta^2-4\alpha\gamma}}{2\alpha}
\end{align}
where we have
\begin{align}
\alpha &= \dfrac{c\dot{a}_j}{2a_j^2} + \frac{1}{2}\left(\frac{(\textbf{r}_j\cdot\textbf{v}_j)^2}{\|\textbf{r}_j\|^3} - \frac{(\textbf{v}_j \cdot \textbf{v}_j)}{\|\textbf{r}_j\|}\right)\\[8pt]
\beta &= -\dfrac{c}{a_j} - \dfrac{(\textbf{r}_j\cdot\textbf{v}_j)}{\|\textbf{r}_j\|} \\[8pt]
\gamma &= \Theta_{j+1} + \dfrac{c\tau}{a_j} - \dfrac{c\dot{a}_j\tau^2}{2a_j^2} - \|\textbf{r}_j\|
\end{align}
We can now solve for $t_\text{em} = t_j + \delta t$ for all particles in any snapshot of the simulation using extrapolation with second order approximations. 

If we compute $\delta t$ in this way for a given particle at snapshot $j$, and the result is that $\delta t > \tau$ (or equivalently $t_\text{em} > t_{j+1}$), then we discard the recovered $\delta t$, since we will almost certainly obtain a better answer if we extrapolate from the next snapshot $j+1$ instead. If $0 < \delta t \leq \tau$, then we save it (an entry for this particle is created in the lightcone output). After doing this for every particle in snapshot $j$, we then advance to snapshot $j+1$ and repeat the process again. Once we have advanced through each snapshots, the lightcone will be filled.

%% file: 2.2_duplication.tex
Extrapolating particle positions from their velocities at snapshot $j$ to find their lightcone-crossing time will work in general, albeit potentially inaccurate because of our approximations. Upon more careful consideration, however, particle extrapolation suffers from the particularly serious ailment of outputting the same particle at two or more times in the final lightcone. 

\begin{figure}[t!]
    \centering
    \includegraphics[width=13cm]{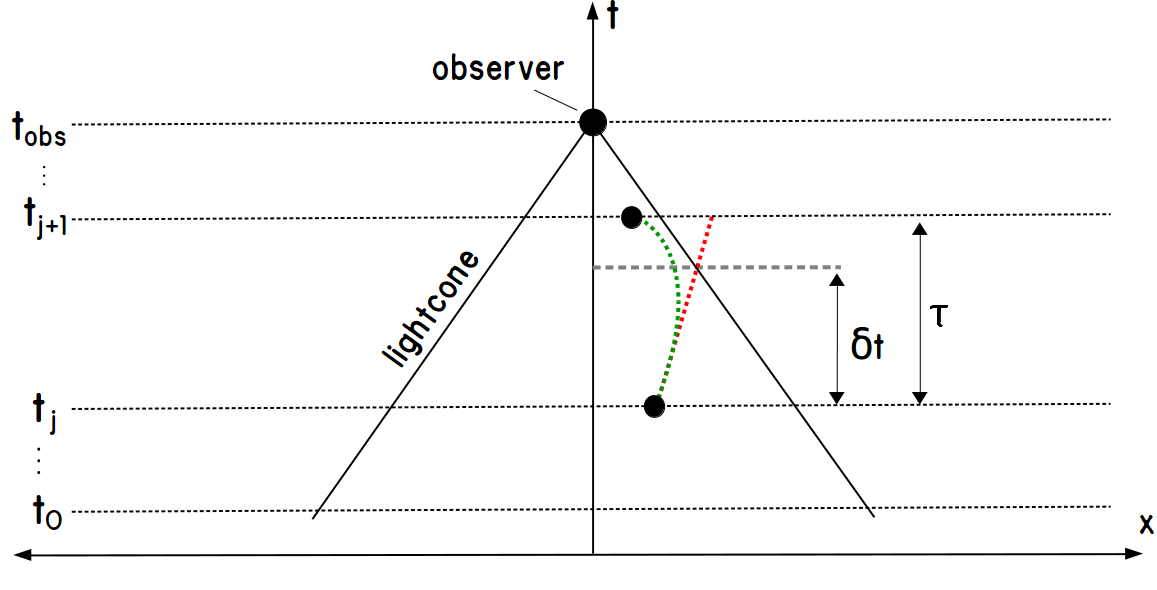}
    \caption{
    \textbf{Figure 2:} A schematic showing the origin of particle duplication issues with the extrapolation approach. The vertical axis is time, and the horizontal is space. Time increases upward. The particle is found at snapshot $j$ with an initial velocity pointing in the $+x$ direction. Extrapolating the position based on this velocity yields the estimated trajectory in red, which crosses the lightcone at time $t_j + \delta t$. The particle's true path is shown in green, which experienced some kind of interaction shortly after the time $t_j$. Since the particle is actually found \textit{inside} the lightcone at time $t_{j+1}$, it must end up crossing the lightcone again, thus being seen by the observer at two different times. The path shapes are greatly exaggerated--- no simulation particles reach such relativistic speeds. 
    } \label{fig:2}
\end{figure}

To understand why this ``particle duplication" happens, consider the diagram shown in \autoref{fig:2}. Here, we have a particle whose velocity \textbf{v}$_j$ at snapshot $j$ suggests it to be moving in the $+x$ direction. Extrapolating it's position places it outside of the lightcone at time $t_{j+1}$ of the next snapshot, and estimates it to have crossed the lightcone at time $t_j + \delta t$. This result is saved and seen in the lightcone output. What \textit{actually} occurred is that the particle underwent some kind of interaction, which is temporally unresolved in the snapshot outputs, that altered it's trajectory. Therefore, it's real path diverges significantly from the extrapolated estimate. In the extreme case as shown in \autoref{fig:2}, the particle won't actually cross the lightcone at all until some later time $t_\text{em} > t_{j+1}$, which will also be saved in the output of some later snapshot. The net effect is that the extrapolation method described in \autoref{sec:extra} places this particle on the lightcone \textit{twice}\footnote{Or even more than twice, if the duplication artifact appears more than once for a particular particle.}, meaning the observer sees the same object at two different redshifts. This of course should be impossible without the object exceeding the speed of light. The symmetric case can also occur, in which certain particles \textit{never} appear in the lightcone.

This duplication error is significant, and effects a few percent of particles, with an even worse impact on halo lightcone catalogs (near 5-10\%). The obvious fix for this is to calculate lightcone-crossing times by particle interpolation rather than extrapolation (in \autoref{fig:2}, a new estimated path connecting the two particle positions would be a much better representation of the true trajectory in green). This approach was not immediately implemented due to the simplicity of the extrapolation method, and lack of foresight into the presently discussed bugs of which we are now well acquainted. Building the lightcone in-situ makes the extrapolation method especially appealing since no knowledge of the upcoming timestep is needed. In any case, particle position interpolation is a necessary requirement for clean and reliable lightcone catalogs, which is described next in \autoref{sec:inter}.

\captionsetup{labelformat=empty}
\captionsetup{labelformat=default}
\vskip-1.5em

%% file: 2.3_interpolation.tex
We have taken a very simple approach to the position-interpolation improvement, so that we would avoid having to perform significant code refactoring (with respect to the original implementation described in \autoref{sec:extra}). Consider again a particle at position \textbf{r}$_j$ (velocity \textbf{v}$_j$) at snapshot $j$, and position \textbf{r}$_{j+1}$ (velocity $\textbf{v}_{j+1}$) at snapshot $j+1$. We first find an ``equivalent linear velocity"--- a constant velocity such that would move the particle from \textbf{r}$_j$ to \textbf{r}$_{j+1}$ in time $\tau=t_{j+1}-t_j$:
\begin{equation}
    \text{\textbf{v}}_\text{lin} = \dfrac{\text{\textbf{r}}_{j+1} - \text{\textbf{r}}_j}{\tau}\>.
\end{equation}
This of course requires that, when working on a particular snapshot $j$ in the process of lightcone construction, we also read in snapshot $j+1$ (not necessary in the original implementation). Once we have done this, the result of Eq.(20) can simply be fed into the framework of \autoref{sec:extra}, where
$\textbf{v}_\text{lin}$ would replace any occurrence of $\textbf{v}_j$. In that case, Eq.(10), which becomes 
\begin{align}
    ||\text{\textbf{r}}|| &= ||\text{\textbf{r}}_j + \text{\textbf{v}}_\text{lin}\delta t||\nonumber\\
    &= \left[(x_j + \dot{x}_\text{lin}\delta t)^2 + (y_j + \dot{y}_\text{lin}\delta t)^2 + (z_j + \dot{z}_\text{linear}\delta t)^2\right]^{1/2}\>,
\end{align}
which is exactly an interpolation. Eq.(13) is then the approximated interpolated position which allows us to solve for $\delta t$, and the rest of the procedure is unchanged. Obviously, by using a linear interpolation, we are enforcing a discontinuous derivative in our particle positions. Really, we could push that discontinuity back to second order (we \textit{could} use the velocities $\textbf{v}_j$ and $\textbf{v}_{j+1}$ to perform a nonlinear interpolation, rather than throwing them away in favor of $\textbf{v}_\text{lin}$). 

In any case, any low-order interpolation (and certainly extrapolation) will inevitably cause some diffusion in the matter distribution of the output lightcone at small scales \cite{2013MNRAS.429..556M}. See \autoref{sec:nonlin} for further discussion on the level of accuracy that LANTERN is able to achieve.

%% file: 3_nonlinear.tex
Through the implementation and testing of the problem solution laid out in \autoref{sec:deriv}-\autoref{sec:inter}, we have discovered several shortcomings, in which our approximations manifest as discreteness effects and other unwanted artifacts in the final lightcone outputs. We should of course like to minimize these effects, and remove them entirely where possible, so it will be useful to go over a road map of all approximations made up until this point, and what their behavior may be in cases of interest. 

Our condition defining the lightcone surface, Eq.(8), is exactly correct. Downstream, however, we dropped higher order derivatives and terms on each side of this equation, which we should consider with care. Here is the expanded form of the Eq.(8) that we derived using discrete simulation outputs, Eq.(18), again:
\begin{align}
    \Theta_{j+1} + \frac{c}{a_j}\left[(\tau-\delta t) - \frac{\dot{a}_j}{a_j}\frac{(\tau^2-\delta t^2)}{2}\right] &= \|\textbf{r}_j\| + \delta t\dfrac{(\textbf{r}_j\cdot\textbf{v}_j)}{\|\textbf{r}_j\|} - \frac{\delta t^2}{2}\frac{(\textbf{r}_j\cdot\textbf{v}_j)^2}{\|\textbf{r}_j\|^3} + \frac{\delta t^2}{2}\frac{(\textbf{v}_j \cdot \textbf{v}_j)}{\|\textbf{r}_j\|}\tag{18}
\end{align}
First, focusing on the left-hand side; the second summand, here, was evaluated as the integral in Eq.(17). During the second line of that equation, we make the approximation that $a(t_j + \delta t) \approx a_j + \dot{a_j}\delta t$, or $\{d^na/dt^n = 0\>:\>n \geq 2\}$. In principle, we could analytically obtain higher order derivatives of $a(t)$ from the Friedmann equation, though the scale factor evolution is quite smooth, and any noticeable impact is doubtful. More importantly, to get the final line of Eq.(17), we perform a first-order Taylor expansion about $a_j$ to evaluate the integral. In doing this, we have again invoked the assumption that $\delta t \ll 1$, which should be acceptable for any simulation with sufficiently fine temporal resolution in its output snapshots. 

Turning our attention to the right-hand side of Eq.(18); as done with the scale factor, we have assumed that $\{d^n\textbf{r}/dt^n = 0\>:\>n \geq 2\}$. For the extrapolation approach, this is the best that we can do, as no acceleration is exposed to us. For the interpolation routine, however, we could make an estimate of $\ddot{r}$ as $(\textbf{v}_{j+1} - \textbf{v}_j) / \tau$, which would relax the restriction that all particle trajectories be represented as linear functions. This may be a desired change to make in the future, and would particularly benefit high density environments; applying simple linear interpolation to particles in tight orbits, e.g. at halo centers, pushes them inward and artificially boosts small-scale power. 

Finally, we must note that, in the code, we do not actually evaluate the traditional quadratic form as is written in Eq.(20). Rather, we perform an approximated quadratic solution, expanded to second order, which is
\begin{align}
\nonumber \delta t &= \dfrac{-\beta\pm\sqrt{\beta^2-4\alpha\gamma}}{2\alpha}\\
&\approx \frac{-\gamma}{\beta} - \frac{\alpha\gamma^2}{\beta^3}
\end{align}
\noindent
We do this to minimize potential numerical instabilities associated with evaluating the full quadratic solution. There are many such possible instabilities, e.g. the case when $4\alpha\gamma$ is small compared to $\beta^2$ (and the sign of $\beta$ and that of the radical term are the same) can give us catastrophic cancellation. One can try to avoid this problem by also using the \textit{Citardauq Formula}, which inverses the sign of the radical term for any particular solution, and choosing to use the more stable expression for the root $\delta t$. We have instead made the more casual decision to use the approximated solution, Eq.(26), which still does eliminate numerical artifacts which otherwise cause the volume of our lightcone output to explode at high redshifts. 

\begin{figure}[]
	\begin{subfigure}{0.5\textwidth}
		\includegraphics[width=\linewidth]{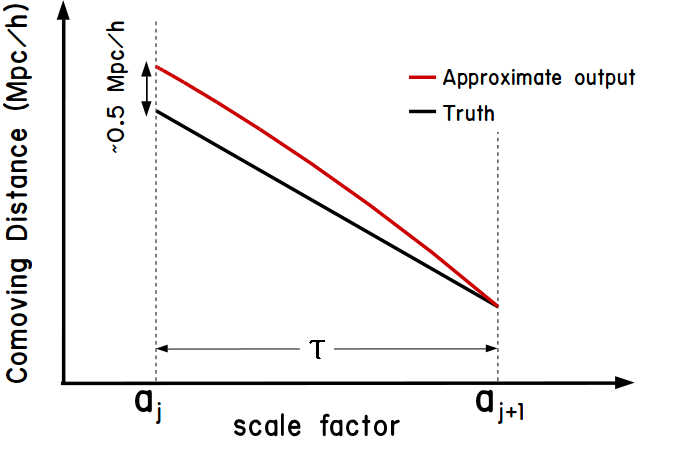}
		\caption{} \label{fig:3a}
		\end{subfigure}
	\hspace*{\fill}
	\begin{subfigure}{0.5\textwidth}
		\includegraphics[width=\linewidth]{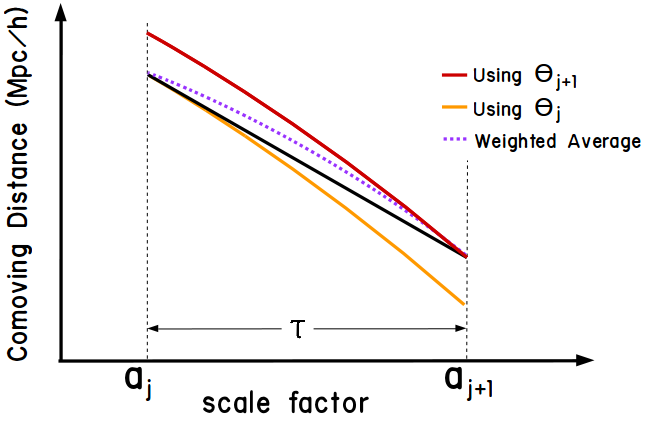}
		\caption{} \label{fig:3b}
	\end{subfigure}
	\hspace*{\fill}
    \caption[Caption for LOF]{
    \textbf{Figure 3:} A toy-figure showing all the qualitative features of the comoving distance-redshift relation, as recovered from the output of the lightcone problem solution described in \autoref{sec:deriv}-\autoref{sec:inter}. \textbf{Fig(3a)}: The output relation if we apply the particle interpolation routine to Eq.(18-23). Due to our approximations, the relation departs from the truth by a maximum of \textasciitilde0.5Mpc at the high-z end. This error repeats for each step $j$, introducing a sawtooth pattern in the relation. \textbf{Fig(3b)}: The same as Fig(3a), with the same relation overplotted, in the case that we replace $\Theta_{j+1}$, in Eq.(18) and Eq.(23), with $\Theta_j$. The error in these two answers appears symmetric, and we can take a weighted average of the two answers to mitigate our approximation errors. Doing this ultimately gives a maximum departure from the true relation of \textasciitilde100kpc.}
    \label{fig:3}
\end{figure}

While we have not dug deep enough to discover the all of the specific influences that each of the above discussed approximations have on our answer, we are aware that compounding all of them has adverse effects. The statistic in which this is most obviously seen is the comoving-distance-redshift relation. We consistently find that our solution over-estimates the distance to the lightcone surface at the high-redshift end of each timestep bounded by snapshots $j$ an $j+1$, as illustrated in the toy plot given in \autoref{fig:3a}. Specifically, we see that particles whose lightcone crossing time was found to be $\delta t \rightarrow 0$ are around \textasciitilde0.5Mpc too far from the true lightcone surface (the black curve in \autoref{fig:3a}, as computed from theory, given the simulation cosmology). Alternatively, we can think of this effect as simply being an apparent increase in the speed of light, in some way due to our approximations.

In response to this issue, we offer what may be seen as a workaround if not a remedy. Notice that the distance-redshift relation recovered from our solution (red curve in \autoref{fig:3a}) is exactly correct for particles that cross the lightcone at $\delta t = \tau$. To understand this, let's look more closely at the LHS of Eq.(18):
\begin{align}
    \int_{a_\text{em}}^1 c\dfrac{da}{a^2H} &= \int_{a_\text{em}}^{a_{j+1}} c\dfrac{da}{a^2H} + \int_{a_{j+1}}^1 c\dfrac{da}{a^2H}\> \nonumber \\[8pt]
    &= \frac{c}{a_j}\left[(\tau-\delta t) - \frac{\dot{a}_j}{a_j}\frac{(\tau^2-\delta t^2)}{2}\right] + \Theta_{j+1} \nonumber
\end{align}

\captionsetup{labelformat=empty}

\noindent
From here, it should be clear that we expect a good answer at $\delta t = \tau$, because in that case, $a_\text{event} = a_{j+1}$, and the integral over $[a_\text{event}, a_{j+1}]$ vanishes. That only leaves the integral represented as $\Theta_{j+1}$ to worry about, which is just a numerical Simpson's method calculation. Simpson's method can practically be as accurate as we'd like it to, with such a smooth function, and all of our approximations in $a(t)$ have dropped out. 

Given this, we see that there is a symmetry to be taken advantage of in this problem. We'd like to calculate the entire integral over $[a_\text{event}, 1]$ numerically, though we cannot trivially do that, since our lower integration bound is unknown. We therefore made the arbitrary choice to split this integral into $[a_\text{event}, a_{j+1}]$ and $[a_{j+1}, 1]$ pieces, though we could have just as well come up with the following arrangement:
\begin{align}
    \int_{a_\text{em}}^1 c\dfrac{da}{a^2H} &= \int_{a_j}^1 c\dfrac{da}{a^2H} - \int_{a_j}^{a_\text{em}} c\dfrac{da}{a^2H}\> \nonumber \\[8pt]
    &= \Theta_{j} - \frac{c}{a_j}\left[\delta t - \frac{\dot{a}_j}{a_j}\frac{\delta t^2}{2}\right]\>\>,
\end{align}
where we have introduced $\Theta_j$ as 
\begin{equation}
    \Theta_{j} \equiv \text{Simpson}\left(\int_{a_{j}}^1 c\dfrac{da}{a^2H}\right)\>.
\end{equation}

Re-deriving the expression for $\delta t$ using Eq.(27) leads to the exact solution as given in Eq.(19-23), except that the constant term in the quadratic has it's summands involving $\tau$ removed, and $\Theta_{j+1}$ is replaced with $\Theta_j$:
\begin{equation}
\gamma' = \Theta_{j} - \|\textbf{r}_j\|\>\>.
\end{equation}
If we re-solve for $\delta t$ (Eq.(26)) using this new constant quadratic term, then we should now have perfect agreement with the true distance-reshift relation at \textit{high}-z, rather than low, with our approximations now manifesting as error at \textit{low}-z. This is shown in \autoref{fig:3b}, and we will refer to the result of solving for $\delta t$ in this revised calculation as $\delta t'$. 

Now that we have established a way to move the approximation-induced error to opposite ends of the time domain of a simulation step, we can leverage this symmetry by using a weighted average of our two measures of the lightcone crossing time per particle, $\delta t$ and $\delta t'$. As should be clear from the curves in \autoref{fig:3b}, the weights for this average should be something like $\delta t_\text{true} / \tau$. Since we do not have $\delta t_\text{true}$, and have no real way to assert the accuracy of $\delta t$ against $\delta t'$, or vice-verse, we will use the mean value of these two vales in the weighting:
\begin{equation}
w = \frac{\delta t + \delta t'}{2\tau}
\end{equation}
with our improved lightcone crossing time given as 
\begin{equation}
\delta tw + \delta t'(1-w)
\end{equation}
This procedure results in the \textasciitilde0.5Mpc/h error in the distance-redshift relation at the earlier timestep endpoint, mentioned above, being reduced to a maximum of \textasciitilde100kpc at the center of the timestep (the new error is the dashed purple curve in \autoref{fig:3b}, which is still not quite zero, since the concavity of each solution has the same sign). 

This workaround, while clever, is perhaps cumbersome and inelegant. The ideal way to deal with nonlinear effects is to take a fully numerical approach. That is, \textit{both} integrals of Eq.(14) should be solved iteratively, rather than one of them being given an approximate analytical evaluation. In other words, the second summand on the LHS of Eq.(18) should also be a numerical evaluation. This seemingly obvious task is nontrivial due to the fact that one of the integration limits itself is bound up in the RHS of Eq.(18), though some sort of root-finding approach should be possible.

%% file: 4.2_geometry.tex
An important characteristic of numerical simulations that have guided our problem-solving thus far is that information is only known at discrete time snapshots. Our approach, conclusions, and notation all acknowledge this limitation. However, we have, thus far, implicitly assumed that we have infinite spatial information, or at least that we have spatial information out to the furthest possible extent of our lightcone. Of course, this is not true; the largest HACC simulations volumes that have been run are expanding cubes whose comoving side-lengths are in the neighborhood of 4Gpc/h. If we place an observer at the corner of one such simulation volume, to what extent can our lightcone reach? The lightcone surface is a spherical object centered on the observer, and it's radius could not exceed, in this case, a comoving distance 4Gpc/h--- a distance which light will cross in a reasonable amount of time, and, assuming a WMAP7 cosmology, only allows our observer to see to a maximum redshift of \textasciitilde2.3. 

On top of this distance limitation, we have also introduced a direction limitation; situating our observer in the corner of the box maximizes the radius of the lightcone, but only allows one octant of the sky to be seen. Placing an observer, instead, in the center of the box allows for a full-sky lightcone to be generated, but in that case the lightcone extent would be reduced to redshift \textasciitilde0.8. These issues are of course only more constraining with smaller simulation volumes.

\begin{figure}[]
	\begin{subfigure}{0.4\textwidth}
		\includegraphics[width=\linewidth]{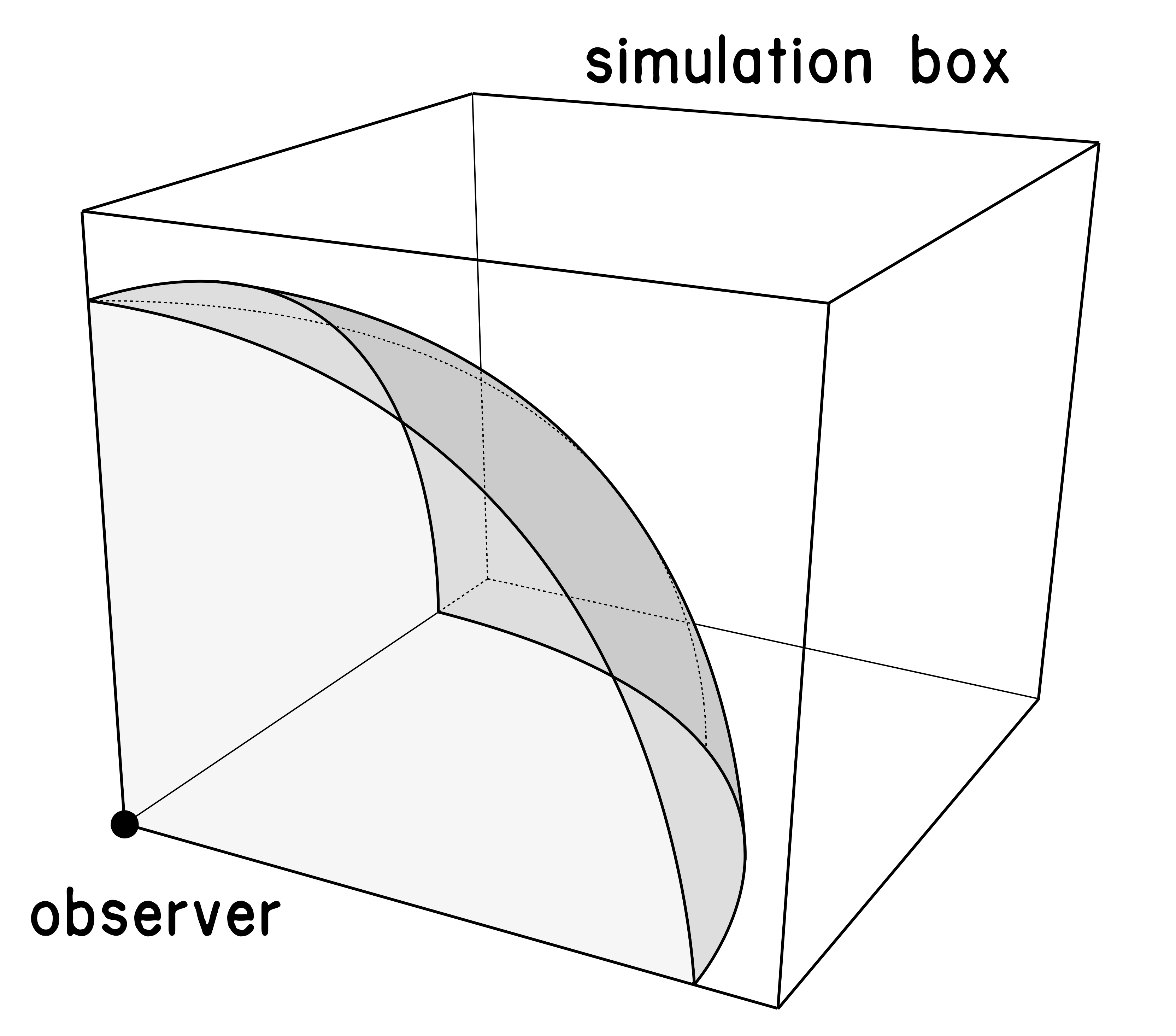}
		\caption{} \label{fig:4a}
		\end{subfigure}
	\hspace*{\fill}
	\begin{subfigure}{0.6\textwidth}
		\includegraphics[width=\linewidth]{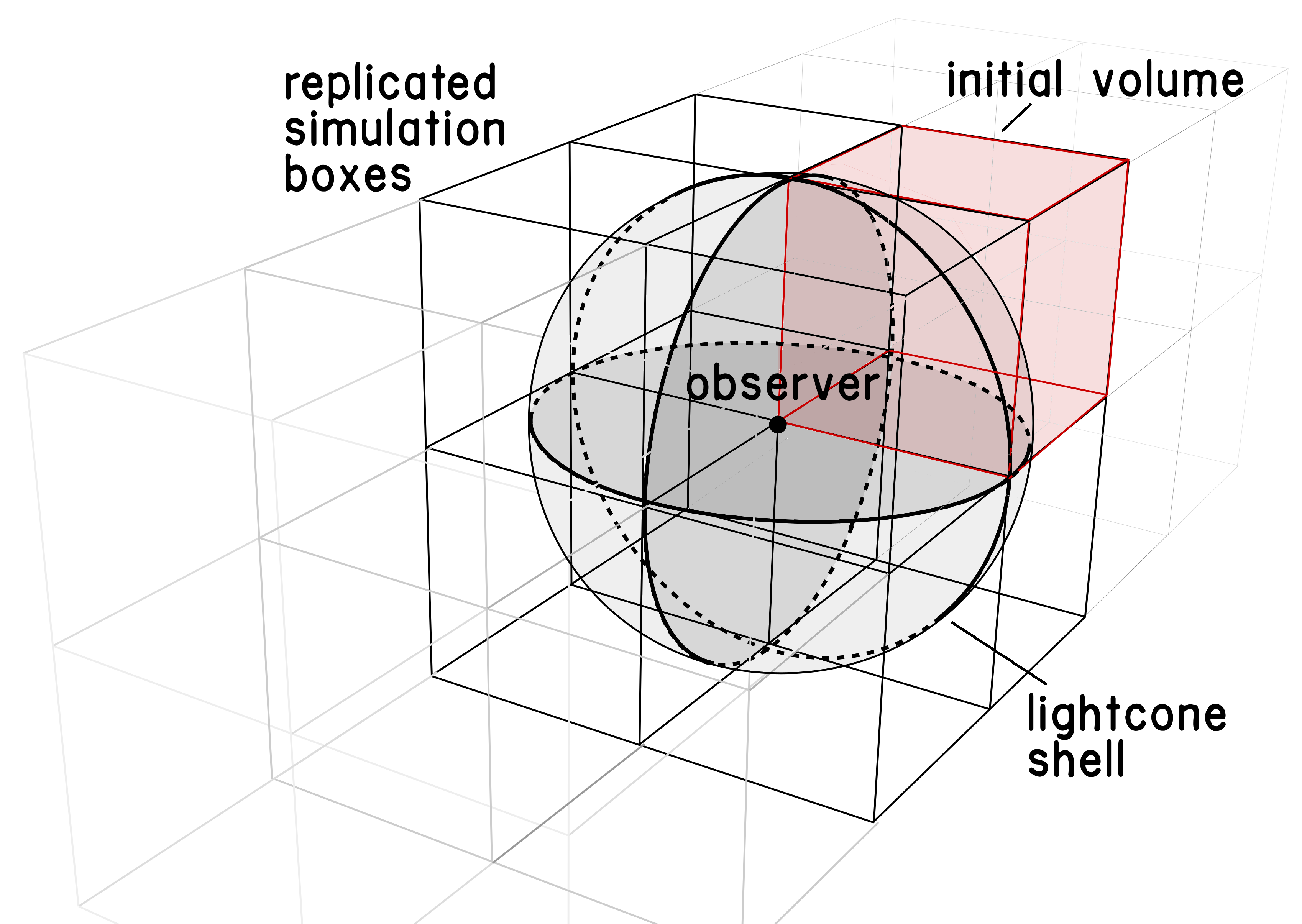}
		\caption{} \label{fig:4b}
	\end{subfigure}
	\hspace*{\fill}
    \caption[Caption for LOF]{
    \textbf{Figure 4:} A comparison of the problem geometry in the case that one simulation volume is used, with the case that the volume is replicated and tiled. \textbf{Fig(4a)}: A view of a simulation volume containing a lightcone, filling an octant of the sky, being generated with respect to an observer placed at the origin. The spherical shell is the extent of the lightcone at some time $t$. Since the extent of the lightcone has not reached the other side of the box yet (light from the other side of the box has yet to reach the observer), it is true that $t < L/c$, where $L$ is the comoving box length. Notably, we \textit{cannot} generate a lightcone, using this simulation volume, to allow the observer to see events occurring beyond a time $t_{\text{max}} > L/c$ into the past. \textbf{Fig(4b)}: A view of a full-sky lightcone now being generated across a tiling of duplicated simulation boxes. The red volume indicates the initial volume shown in Fig(4a). Notice that now we have eight times the sky-coverage, without having had to reduce the radius of the lightcone. We can replicate the simulation boxes as many times as we'd like to produce an arbitrarily large lightcone (albiet with the usual consequences of finite-volume effects associated with resampling the same structure).
    }
    \label{fig:4}
\end{figure}

Surely we would like to fill lightcones with simulation particles/objects to any arbitrary redshift we choose, and surely we would also like the option to do so over the entire angular domain of any observer. Our solution implemented in LANTERN is to grow the spatial extent of our simulation output, effectively, by replicating the box many times and ``tiling" these replications in space. This idea is shown visually in \autoref{fig:4}. For example, if we'd like to fill an all-sky lightcone out to a redshift of 2 (a comoving distance of \textasciitilde3.7Mpc/h given WMAP7) with particle output from a 2Gpc/h simulation, we would replicate the box once in each Cartesian direction for a limiting lightcone radius of 4Gpc/h.

While implementing this replication technique does solve the issues described above, and allows us to generate a lightcone to any arbitrary redshift, it introduces some further complications of its own. If we carelessly replicate the simulation volume many times, the presence of repeating large-scale structure will start to become visually obvious, and inject artifacts into various statistics, e.g. power spectra, redshift distributions, etc. For this reason, the code has been given the capability to randomly rotate each replication by swapping the $x$, $y$, and/or $z$ positions and velocities of its particles.
This solution, in turn, has yet another side-effect: before rotating the replicated boxes, they are all seamlessly joined to each other due to the periodic boundary condition of each box edge. This property will be lost after randomly-rotating each volume, and discontinuities can be created at box edges (filaments and halos will be clipped).

\begin{figure}[]
  \begin{minipage}[c]{0.65\textwidth}
    \includegraphics[width=\linewidth]{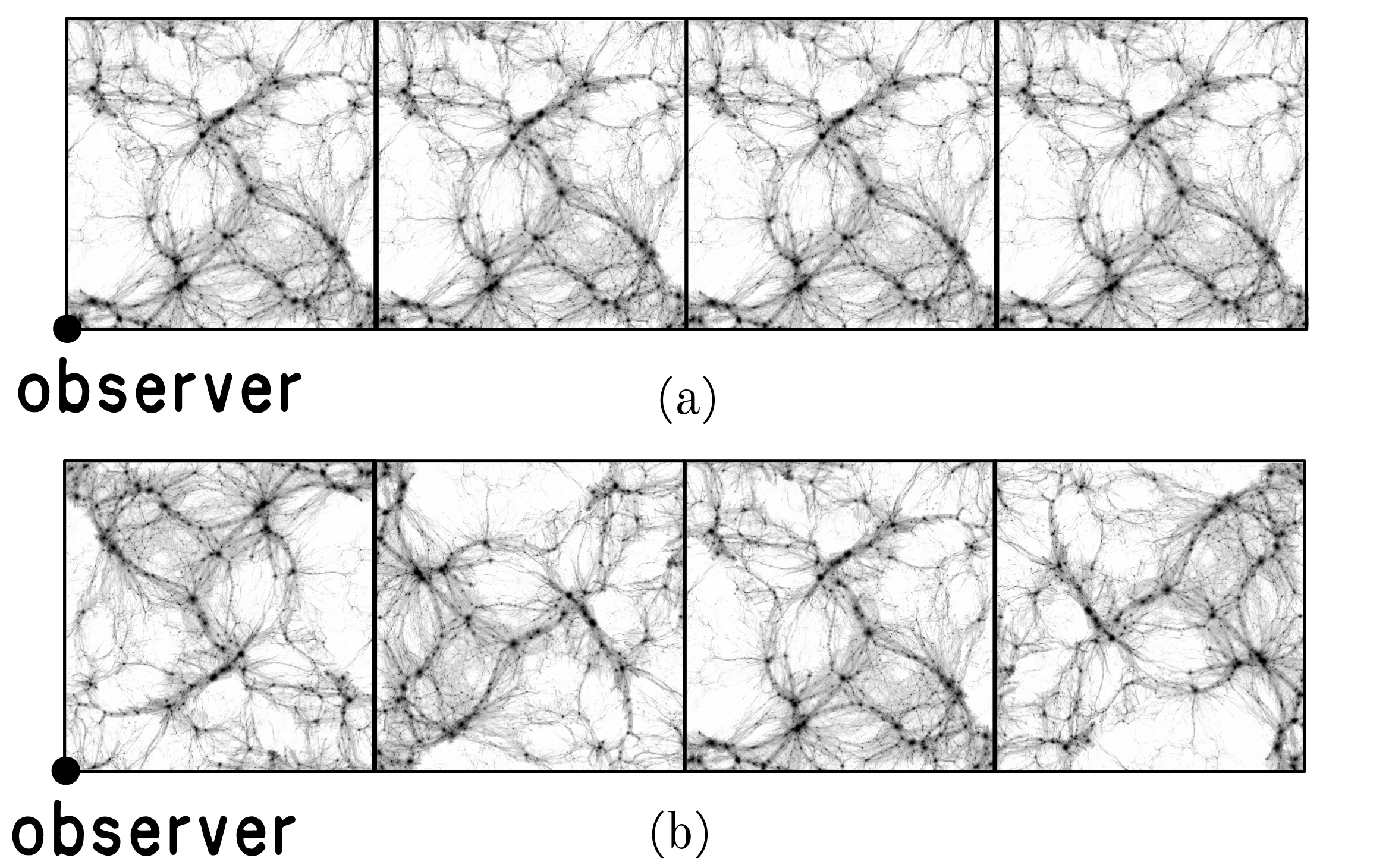}
  \end{minipage}\hfill
  \begin{minipage}[c]{0.35\textwidth}
	\caption[Caption for LOF]{\textbf{Figure 5:}\footnotemark A comparison of the consequences of simulation box replications, with (Fig(5b)) and without (Fig(5a)) enabling random rotations. The problem of repeating LSS is clear in Fig(5a), as the density along the lower box edge is noticeably higher than the top edge. Repeating structures along this lower edge have very little angular separation in the observer's line of sight, making the effect particularly strong. These issues are cured in Fig(5b), though some density discontinuities are obvious (notably along the vertical edge separating the initial and first replicated volumes).}
	\label{fig:5}  
  \end{minipage}  
\end{figure}

In short, we are forced to decide between introducing either repeating large-scale structure, or large-scale edge effects, into the lightcone output. The latter has tended to be the less significant problem. These issues are further detailed in \autoref{fig:5}, where the box replications along a single axis are visualized.

This replication and rotation approach has been implemented in the production of previous simulated skies \cite{Blaizot:2003av, Kitzbichler:2006ec, bernyk2016}. Though this strategy does decorrelate particles at the replicated box boundaries, and should in principle manfiest as an artifact in the correlation function, we find it negligible in practice when running LANTERN on very large simulation boxes. Smaller boxes require more frequent replication, which will worsen the effect, but then again would also worsen the projection effects (or ``kaleidoscope" effects) in the event that we did not rotate. With the increasingly large HACC N-body simulations that LANTERN is typically run, this is all mostly a non-issue. Still, for a quantitative investigation into the impact of this method, the reader is referred to \cite{Blaizot:2003av}.

%
%

\captionsetup{labelformat=empty}

%
%
 
\footnotetext{Credit for the cosmic web image used in the creation of this figure to Mansfield \& Diemer: http://www.benediktdiemer.com/visualization/images/}

\captionsetup{labelformat=default}


%% file: 5_export.tex
 A brief note on package exportability; LANTERN is currently woven into the HACC codebase, and is not modular. It is intended to export it as a standalone tool which will interface with simulation snapshots, though it will likely always expect particles written in HACC's custom data format GenericIO\cite{2016NewA...42...49H}, and need to call HACC utilities for computing cosmological quantities. Thus, we can only suggest that any interested readers attempt to implement the solution presented here independently; please send any questions/comments to \texttt{jphollowed@anl.gov}.

%% file: 6_credits.tex
The work described over the previous pages was conducted at Argonne National Laboratory, Cosmological Physics and Advanced Computing group. The original extrapolation code implementation was done by Hal Finkel, upon which our work has been built. Significant extensions and improvements were completed by Joe Hollowed beginning in December 2017, with indispensable collaboration with Patricia Larsen and Steve Rangel. Guidance throughout the project was given by Katrin Heitmann and Salman Habib, as well as valuable input from other members of CPAC; Lindsey Bleem, JD Emberson, Danila Korytov, and Nesar Ramachandra. Our laughably (if not offensively) forced acronym, LANTERN, is thanks to \cite{2019arXiv190312180C}.